
\input phyzzx
\nopagenumbers

\def\fun#1#2{\lower3.6pt\vbox{\baselineskip0pt\lineskip.9pt
  \ialign{$\mathsurround=0pt#1\hfil##\hfil$\crcr#2\crcr\sim\crcr}}}
\def\lap{\mathrel{\mathpalette\fun <}}
\def\gap{\mathrel{\mathpalette\fun >}}
\def\order{{\cal O}}
\def\etal{{\it et al.}}

\def\rar{\rightarrow}
\def\Rf{\normalbaselines\parindent=0pt \medskip\hangindent=3pc \hangafter=1 }

\Pubnum={IASSNS-HEP-93/21\cr IASSNS-AST 93/20}
\date={May 1993}
\titlepage
\title{THE RATE OF THE PROTON-PROTON REACTION}
\author{Marc Kamionkowski\foot{e-mail: kamion@guinness.ias.edu.}\foot{SSC
Fellow} and John N. Bahcall\foot{e-mail: bahcall@guinness.ias.edu}}
\address{School of Natural Sciences, Institute for Advanced
Study, Princeton, NJ 08540}
\abstract
\singlespace
We re-evaluate the nuclear matrix element for the proton-proton
reaction\break $p+p\rightarrow {}^2{\rm H}+ e^+ + \nu_e$, which is
important for stellar-evolution calculations referring to stars
with masses $\lap1\,M_\odot$ and for the solar-neutrino problem.
We self-consistently determine the effect of vacuum
polarization on the matrix element by first correcting the
low-energy scattering data to account for vacuum polarization.
We then calculate the proton-proton wave function by integrating
the Schrodinger equation with vacuum polarization included.
We use improved data for
proton-proton scattering and for the deuteron wave function.  We
evaluate the uncertainties that are due to experimental error and
estimate those that are due to theoretical inadequacies.
Without vacuum polarization, we find that the square of the overlap
integral is $\Lambda^2=6.96$ with an uncertainty of 0.2\% due to errors
in the experimental parameters and an uncertainty of 1\% due to
lack of knowledge of the shape of the nuclear potentials.  We
estimate the theoretical uncertainty by using
six different deuteron potentials and five different
proton-proton potentials.  Vacuum polarization decreases the
calculated value by $0.6_{-0.4}^{+0.1}$\%.  The complete result is
$\Lambda^2=6.92\times(1\pm0.002^{+0.014}_{-0.009})$ where the
first uncertainty is due to experimental errors and the second
uncertainty is due to theoretical uncertainties.  Our value of
$\Lambda^2$ is 2\% smaller than the value obtained in 1969 by
Bahcall and May.  The improved calculations of the rate
of the $pp$ reaction described here increase slightly the
predicted event rates for the chlorine and the
Kamiokande solar-neutrino experiments.

\endpage
\baselineskip=18pt
\overfullrule=0pt
\pagenumber=2
\pagenumbers

\FIG\potentials{Proton-Proton potentials.  The solid curve is the
exponential potential, the dot--short-dash curve is the Yukawa
potential, the short-dash curve is the Gaussian potential,
the long-dash curve is the square-well potential, and the
dot--long-dash curve is the repulsive-core potential.  All five
potentials result in a scattering length $a_p=-7.8196$ and an
effective range $\rho_p=2.790$.}
\FIG\deutplot{Deuteron wave functions.  The solid curve is the
wave function for the SSC potential, the short-dash--long-dash
curve is that for the Urbana v14 potential, the short-dash curve
is that for the Argonne v18 potential, the long-dash curve is
that for the Argonne v14 potential, the dot--short-dash curve is
that for the RSC potential, and the dot--long-dash curve is the
McGee wave function.}
\FIG\overlap{Overlap of $pp$ and deuteron wave functions.  As in
Fig.~1, the solid curve comes from using the exponential $pp$
potential, the dot--short-dash curve from the Yukawa potential, the
short-dash curve from the Gaussian potential, the long-dash
curve from the square-well potential, and the dot--long-dash
curve from the repulsive-core potential.  In all five cases, the
Argonne v14 deuteron wave function is used.  In (a) we show the
overlap out to a radius of 50 fm, while in (b) we magnify the
first 5 fm.}

\unnumberedchapters

\centerline{1. INTRODUCTION}
\medskip

The cross section for the basic reaction that initiates the proton-proton
fusion chain, $p+p\rar {}^2{\rm H} + e^+ + \nu_e$, is important for
calculations of the stellar evolution of main-sequence stars of solar
mass or less.  In particular, the calculated flux of the crucial ${}^8{\rm
B}$ solar neutrinos depends approximately upon the -2.6th power of the
square of the $pp$ matrix element (Bahcall and Ulrich 1988).
The rate for the $pp$ reaction was first estimated by
Bethe and Critchfield (1938), and then calculated with an
effective-range approximation by Salpeter (1952).  The results
of detailed numerical calculations were presented by Bahcall and
May (1969); various specific calculations and corrections have been
published since then (Brolley 1971; Gari and Huffman 1972; Gari
1978; Bargholtz 1979; Guessoum 1988; Gould and Guessoum 1990;
and Carlson \etal\ 1991).

In this paper we re-calculate the nuclear
matrix element for the $pp$ reaction using improved data and an
explicit and consistent treatment of vacuum-polarization
effects.  We also identify and evaluate the various sources of
uncertainty.

The low-energy cross-section factor for the $pp$ reaction can be
written (see Bahcall \etal, 1982)
$$
S_{pp}(0)=3.89\left({\Lambda^2\over 6.92}\right) \left( {G_A/G_V
\over 1.2573} \right)^2 \left({1+\delta \over 1.01}
\right)^2\,\times\, 10^{-25}\, {\rm MeV-barns}.
\eqn\Szero
$$
The value of $G_A/G_V=1.2573(28)$ (Hikasa \etal\ 1992) used here
represents a weighted average of five precise modern
experiments in which the correlation between the electron
momentum and the neutron spin or the proton recoil spectrum are
measured. The quantity $\delta=0.01^{+0.02}_{-0.006}$ (Bahcall
and Pinsonneault 1992) is
the fractional correction to the nuclear matrix element due to
the exchange of $\pi$ and $\rho$ mesons.  We do not re-evaluate
$\delta$ in this paper (for a discussion of the mesonic
corrections, see, \eg, Blin-Stoyle and Papageorgiou 1965, Gari and
Huffman 1972, Dautry, Rho, and Riska 1976, Bargholtz 1979, and
Carlson \etal\ 1991).  We use our current result,
$\Lambda^2=6.92$ for the square of the integral of the overlap
of the proton and deuteron wave functions, which is about 2\%
smaller than the previous best estimate $\Lambda^2=7.08$.  Small
effects due to electromagnetic
radiative corrections are taken into account by using $ft$
values for nuclear beta decays which have {\it not} been altered
by theoretical corrections for radiative processes (Bahcall and
May 1968).

We focus on
$\Lambda^2$, the square of the overlap integral
(at zero energy) where (Salpeter 1952; Bahcall and
May 1969):
$$
\Lambda=\left( {a_p^2 \gamma^3 \over 2} \right)^{1/2} \int
u_d(r) u_{pp}(r)\,dr.
\eqn\defnoflambda
$$
The quantity $a_p$ is the $pp$ scattering length, and $\gamma=(2\mu
E_d)^{1/2}$ is the deuteron binding wave
number (where $\mu$ is the proton-neutron reduced mass and $E_d$
is the deuteron binding energy).  The function $u_{pp}(r)$ is
the radial part of the initial $pp$ wave function
and $u_d(r)$ is the radial part of the $S$-state deuteron wave
function.

In the most systematic previous study, Bahcall and May
(1969) carried out calculations for a number of
different nuclear potentials and wave functions and found
$\Lambda^2=7.08(1\pm0.025)$.  Since 1969, there have
been several
published calculations of $\Lambda^2$ (summarized in Table II of
Bahcall and Pinsonneault 1992) with specific assumptions about
the nuclear potentials; the values obtained in all but one case
ranged from 6.83 to 7.04.  Only
the calculation by Gould and Guessoum (1990; Guessoum 1988)
gives a larger value ($\Lambda^2=7.39$) than the best estimate
of Bahcall and May.  (We discuss problems with the
Gould-Guessoum calculation at the end of Section 5.)  In some of
the previous calculations, the effect of vacuum polarization was
included or partially included, but was
not isolated, which makes comparison between the different
calculations difficult.

We first calculate $\Lambda^2$---without including vacuum-polarization
effects---using improved data for $pp$ scattering and for the deuteron
wave function.  We investigate systematically the uncertainties
in our result, which are caused in part by errors in the
experimental measurements of various input parameters and in
part by the imprecisely known shapes of the $pp$ and $np$ nuclear
potentials.  The theoretical uncertainty is
smaller than is suggested by a naive comparison of the different
values for $\Lambda^2$ obtained using the nuclear
potentials available in the literature.
Each potential in the literature predicts slightly
different values for the measured quantities.  If we constrain
each nuclear potential to reproduce the same values for the
measured parameters, the apparent differences in the
calculated values of $\Lambda^2$ are reduced.  We find
$\Lambda^2_{\rm no~vac.~pol.}=6.96$.  The
uncertainty due to experimental errors is about 0.2\% and that
due to theoretical uncertainties is about 1.0\%.

We then include the effect of vacuum polarization in a
self-consistent way by correcting the nuclear parameters
derived from low-energy $pp$ scattering data and by
including vacuum polarization in the Schrodinger equation used to
calculate the $pp$ wave function.
The first effect raises the
value of $\Lambda^2$ by $0.4^{+0.4}_{-0.1}$\% and the second
correction
decreases $\Lambda^2$ by about 1.0\%.  Both effects must be
included in a self-consistent treatment.  Therefore, vacuum
polarization decreases the calculated value of $\Lambda^2$
by about $0.6^{+0.1}_{-0.4}$\%.  The final result is
$$
\Lambda^2=6.92\times(1\pm0.002^{+0.014}_{-0.009}),
\eqn\theanswer
$$
where the first uncertainty is due to experimental errors and the
second is due to uncertainties in the nuclear potential.

In Section 2, we discuss the effective-range approximation and
use it to estimate the uncertainty in $\Lambda^2$ due to the
errors in the input parameters that are measured.  In Section 3, we
discuss the $pp$ wave functions, and in Section 4, we discuss
the deuteron wave functions.  Our numerical results are
presented in Section 5.  In Section 6, we discuss the
effect of including vacuum polarization in our analysis, and in
Section 7, we summarize our results and discuss briefly
implications for predicted solar-neutrino rates.

\bigskip
\centerline{2. EFFECTIVE-RANGE APPROXIMATION}
\medskip

Before presenting our numerical results for $\Lambda^2$ it is
useful to recall the calculation of $\Lambda^2$
based on the effective-range approximation (Salpeter 1952;
Ellis and Bahcall 1968; Bahcall and May 1969).

If $\rho_p$ and $\rho_d$ are the $pp$
and deuteron effective ranges, $a_p$ is the proton scattering
length, and $\gamma$ is the deuteron binding wave
number, then
$$
\Lambda_{eff}(0)=\widetilde\Lambda(0) + \left({\gamma^2 a_p N \over 4 }
\right) (\rho_d + \rho_p),
\eqn\effectiverange
$$
where
$$
\widetilde\Lambda(0) = N\left\{1+ {a_p \over R}\left[ E_1(\chi)
- {e^{-\chi} \over \chi} \right] \right\} e^\chi,
\eqn\twiddle
$$
and $N=[(1+\eta_d^2)(1-\gamma\rho_d)]^{-1/2}$.  Here,
$\eta_d$ is the asymptotic ratio of
$D$- to $S$-state deuteron wave functions, and $\chi=(\gamma
R)^{-1}$.

The low-energy scattering parameters $a_p$ and $\rho_p$ that are
used in Eqs.~\effectiverange\ and \twiddle\ are those determined
by fitting low-energy $pp$ phase shifts measured relative to
a pure Coulomb potential [see Eq.~(11) below].  Published
values of $a_p$ and $\rho_p$ that appear in the literature are
often corrected for various electromagnetic and/or
strong-interaction effects.  These corrected values of $a_p$ and
$\rho_p$ cannot be used here.  Using
$a_p=-7.8196(26)$ fm, $\rho_p=2.790(14)$ fm (Bergervoet \etal,
1988), $E_d=2.224575(9)$ MeV, $\eta_d=0.0256(4)$, and
$\rho_d=1.759(5)$ (Brandenburg \etal, 1988), we find
$$
\Lambda_{eff}^2(0)= 6.975\times \left[
1-0.29\,(\rho_d-1.759)-0.10\,(\rho_p-2.790)\right].
\eqn\effanswer
$$
We have exhibited in Eq.~\effanswer\ the dependence of the
result on the effective ranges, which are the experimental
quantities with the largest
uncertainties.  The uncertainty in $\Lambda^2$ due to the
errors in the experimentally determined quantities is
0.2\% and comes predominantly---but not exclusively---from the
imperfect experimental knowledge of
$\rho_d$ and $\rho_p$.

The effective-range approximation yields a value for
$\Lambda^2$ consistent with what is obtained
numerically  by integrating the Schrodinger equation ($\sim 6.96$; see
below and Bahcall and May 1969).  The analytic result given in
Eq.~\effanswer\ does not assume any specific knowledge of the
shape of the nuclear potential, reflecting the fact that
most of the overlap between the proton and the deuteron wave
functions occurs at radii large compared with the range of the
nuclear forces.  Since the wave functions at asymptotically
large radii are accurately determined by the experimentally
determined quantities $\gamma$, $\eta_d$, $\rho_d$, $\rho_p$,
and $a_p$ (Bahcall and May, 1969), $\Lambda^2$ is to a large
extent determined by the experimental quantities and
is therefore insensitive to the details of the nuclear
interaction.  This robustness will be quantified further in the
following sections (see especially the last columns of Table 1
and Table 2).

Since the effective-range approximation gives an answer
in agreement with the numerical integrations, we assume
that the effect of small changes in the experimental parameters
on $\Lambda^2$ may be estimated well by the
effective-range approximation.  This assumption has been
verified quantitatively by comparing changes in the calculated
$\Lambda^2$ produced by varying input parameters in simple
wave functions.  We conclude that the
experimental uncertainty in $\Lambda^2$ is indeed about 0.2\%.

\bigskip
\centerline{3. PROTON-PROTON WAVE FUNCTION}
\medskip

We begin by summarizing the theory of low-energy
$pp$ scattering without including vacuum polarization.  The $s$-wave $pp$
radial wave function $u_{pp}(r)$ satisfies the radial Schrodinger
equation,
$$
{d^2u\over dr^2}-\left[{1\over Rr}+V(r)\right]u = -k^2u,
\eqn\schrod
$$
where $R=\hbar^2/Me^2=28.8198$~fm, $M$ is the proton mass,
$V(r)$ is the nuclear potential multiplied by $M/\hbar^2$,
and $k=Mv/2\hbar$ is the center-of-mass momentum where $v$ is
the relative velocity.  The first boundary condition is that
$u_{pp}(0)=0$ and the second condition is that $u_{pp}(r)$
approaches a properly normalized distorted Coulomb wave at large
$r$.

Since $V(r)$ has finite range, $u_{pp}(r)$ approaches an asymptotic limit,
$\phi(r)$, for large radii.  In this domain, $\phi(r)$ satisfies
$$
{d^2\phi\over dr^2}-\left[{1\over Rr}\right]\phi = -k^2\phi.
\eqn\asschrod
$$
This equation is solved by
$$
\phi(r)=C_0[G_0(kr)+\cot\delta F_0(kr)],
\eqn\phieqn
$$
where $G_0$ and $F_0$ are the irregular and regular Coulomb wave
functions, and $\delta$ is a phase shift.  The Gamow penetration
factor is
$$
C_0^2={2\pi \eta \over \exp(2\pi\eta)-1 },
\eqn\Coeqn
$$
where $\eta=e^2/\hbar v$, and $v$ is the relative velocity of
the two protons.
Note that the normalization of $\phi$ is such that $\phi(0)=1$.
For a given potential, $V(r)$, the phase shifts for
each energy, $\delta(k)$, are calculated by setting $u_{pp}(0)=0$,
numerically integrating Eq.~\schrod\ well beyond the range of
$V(r)$, and then matching the solution onto Eq.~\phieqn.

The scattering length and effective range are determined
for the measured phase shifts in low-energy $pp$
scattering experiments.  The scattering
length, $a_p$, and effective range, $\rho_p$, are related to the
phase shifts by
$$
C_0^2 k\cot\delta + {1\over R} h(kR) = -{1\over a_p}
+{1\over2}\rho_p k^2 + \cdots,
\eqn\expansion
$$
where $h(x)=(x^2/3)[1+(x^2/10)+...]$ for small $x$.
Since the scattering length is given by the limit
$$
-{1\over a_p} = \lim_{k\rightarrow 0} C_0^2 k \cot \delta,
\eqn\apdefined
$$
only the $k\rightarrow0$ solutions to Eqs.~\schrod\ and
\asschrod\ are needed to determine $a_p$.  For $k=0$,
Eq.~\asschrod\ is solved by modified Bessel functions, and one
finds (using the definition of $a_p$) that the $k=0$ limit of
Eq.~\phieqn\ is
$$
\phi(r)= y_1(r/R) - {R\over a_p} y_2(r/R),
\eqn\phikiszero
$$
where
$$
y_1(x)\equiv 2\sqrt{x}K_1(2\sqrt{x}),\quad {\rm and}\quad
y_2(x)\equiv \sqrt(x)I_1(2\sqrt{x}).
$$
Given $V(r)$, the scattering length is then given by evaluating
$$
{1\over a_p} = {2\over R} {r K_0(2\sqrt{r/R})+\alpha\sqrt{rR}
K_1(2\sqrt{r/R}) \over \alpha\sqrt{rR}I_1(2\sqrt{r/R})-r
I_0(2\sqrt{r/R}) },
\eqn\apeqn
$$
at some $r$ large compared with the range of nuclear forces.
Eq.~\apeqn\ converges to $a_p$ as $r$ is increased.  In
practice, a radius $r\gap20$~fm yields a precision better than
that which is determined experimentally.
The logarithmic derivative $\alpha=r u_{pp}'(r)/u_{pp}(r)$ is
determined by solving Eq.~\schrod\ numerically with $u_{pp}(0)=0$.

The effective range $\rho_p$ is defined by the relation
$$
\rho_p=2\int_0^\infty\,[\phi^2(r)-u^2(r)]\,dr,
\eqn\roeqn
$$
and can be determined numerically for a given potential $V(r)$.

To obtain the $pp$ wave function, we need to assume
a form for the nuclear potential {\it that yields the measured
values of $a_p$ and $\rho_p$}, and then integrate the Schrodinger
equation.  We want our estimate of the uncertainties to be conservative;
therefore, we make minimal assumptions about the nuclear potential.
Following the example of Bahcall and May (1969), we
consider several plausible functional forms for $V(r)$.  To fit
the two experimental quantities, $a_p$ and $\rho_p$, each
potential is described by two parameters.

The nuclear interaction is heuristically represented by an
attractive potential of depth $V_0$ and range $b$.  We try the
following forms for the nuclear potential: (i) a square well potential,
$$
V(r)=\cases{V_0 & for $r<b$,\cr 0& for $r>b$;}
\eqn\swpotential
$$
(ii) a Gaussian potential,
$$
V(r)=V_0\exp[-(r/b)^2];
\eqn\gausspotential
$$
(iii) an exponential potential,
$$
V(r)=V_0 \exp(-r/b);
\eqn\exppotential
$$
(iv) a Yukawa potential,
$$
V(r)=V_0 (b/r) \exp(-b/r);
\eqn\yupotential
$$
and (v) a repulsive-core (RC) potential,
$$
V(r)=V_0\exp(-r/b)\, +\, 300\,{\rm MeV}\, \Theta(0.4~{\rm fm}\,-\,r),
\eqn\corepotential
$$
where $\Theta$ is a step function.
Three of these shapes (square well, Yukawa, and exponential)
were used by Bahcall and May (1969).
The values of $V_0$ and $b$ that give $a_p=-7.8196$ fm and
$\rho_p=2.790$ (see Section II) fm are listed in Table I for the five
models, and the potentials are plotted in Fig.~\potentials.

\bigskip
\vfill\eject
\centerline{4. THE DEUTERON WAVE FUNCTION}
\medskip

The deuteron wave function is determined by a procedure
similar to that used to calculate the $pp$ wave
functions.  To estimate the
sensitivity of $\Lambda^2$ to details of the neutron-proton
interaction, we use six wave functions
which fit the static deuteron parameters (\eg, magnetic and
electric quadrupole moments), as well as those
which are more crucial to the calculation of
$\Lambda^2$: $E_d$, $\eta_d$, and $\rho_d$,
[see Eqs.~\effectiverange\ and \twiddle, and the following
discussion].  We are grateful to R. Wiringa for supplying a code
that calculates the deuteron wave functions.
The wave functions are obtained from the Argonne v14 potential (Wiringa
\etal, 1984), the Reid soft-core
(RSC) potential (Reid, 1968), the Urbana v14 potential (Lagaris
and Pandharipande, 1981), the super-soft-core (SSC) potential
(de Tourreil and Sprung, 1973), and the Argonne v18 potential
(Wiringa, 1993); in addition, we also calculate $\Lambda^2$ with
the McGee wave function (McGee, 1966), which was used by Bahcall
and May.

The six wave functions are plotted in
Fig.~\deutplot, and the values of $E_d$, $\eta_d$ and $\rho_d$
are given in Table 2.  Since the potentials used to calculate the wave
functions
were constructed by assuming different values
for the deuteron properties, none of the potentials
precisely reproduces the current values of all of the
experimental quantities of relevance here.  Therefore, the
uncertainties in the wave functions due to uncertainties in the
details of the nuclear interaction are not as large as suggested
by the (already small) differences in the wave functions plotted in
Fig.~\deutplot.  For example, the asymptotic behavior of the
wave function is given by $u_d(r)\rightarrow
N(2\gamma)^{1/2}\exp(-\gamma r)$, where $N$ and $\gamma$ are
fixed by $E_d$, $\rho_d$ and $\eta_d$.  Since these primary quantities
are determined accurately by the existing experimental data [see
discussion following Eqs.~\effectiverange\ and \twiddle], the
behavior of the wave function at large radii is not as uncertain
as suggested by Fig.~\deutplot.  Most
of the overlap integral occurs at large radii.  The value of
$\Lambda^2$ that is obtained by numerically integrating the
overlap of $u_d$ and $u_{pp}$, when $u_d$ has an incorrect asymptotic
behavior, is inconsistent with current knowledge.
In order to obtain an accurate value of
$\Lambda^2$ that uses the current experimental data, we make small
corrections with the aid of
the effective-range approximation.  We noted earlier (end of Section II)
that the effective-range approximation reliably
reflects small changes in $\Lambda^2$ due to small changes in the
parameters.  Specifically, we evaluate $\Lambda^2$ for a given deuteron
potential from the relation,
$$
\Lambda^2=\left[{\Lambda_{eff}^2({\rm experimental\,
parameters})
\over \Lambda_{eff}^2(\rm model\,
parameters)} \right]\Lambda_{num}^2\equiv C\Lambda_{num}^2,
\eqn\correction
$$
where $\Lambda_{num}^2$ is the value obtained numerically by calculating
the overlap integrals with the six deuteron wave functions, and the values
of $\Lambda_{eff}$ are obtained from the effective-range equations,
Eqs.~\effectiverange\ and \twiddle.  For the
potentials listed in Table 2, the value of $C$ ranges from 0.979
to 0.996.
By using several different potentials, we probe the
sensitivity of $\Lambda^2$ to details of the nuclear interaction.

\bigskip
\centerline{5. NUMERICAL RESULTS FOR $\Lambda^2$}
\medskip

In Fig.~\overlap, we plot the integrand $u_{pp}(r)u_d(r)$ as a
function of radius for the five assumed $pp$ potentials
(using the Argonne v14 deuteron potential).  This figure shows
that drastic changes in the shape of the $pp$
interaction [see Fig~\potentials] result in relatively small
changes in the value of the integrand, and the difference is
significant only in the region $r\lap5$ fm.  The integrand is
insensitive to the shape of the $pp$ potential for
radii $r\gap5$ fm.  In the region where there is a visible
difference ($r\lap5$ fm), the wave functions are constrained by
the measured effective range
[see Eq.~\roeqn], so those wave functions which are smaller in the
region $r\lap2$ fm are larger in the region $r\gap2$ fm (and
{\it vice versa}).  About 40\% of the integrand comes from the
region $r\lap5$~fm and only about 2\% comes from the region
$r\lap1$~fm.

The values of $\Lambda^2$ obtained
using the five different $pp$ interactions are listed
in Table I.  (Here we used the Argonne v14 deuteron potential.)
They range from $\Lambda^2=6.916$ (using the square-well
potential) to $\Lambda^2=6.979$ (using the Yukawa potential) to
$\Lambda^2=6.939$ (using the strong repulsive core).
The difference in $\Lambda^2$ between the pure exponential potential
and the potential that includes a strong repulsive core is only 0.3\%.
Thus we assign a total uncertainty of $\pm0.5\%$ to the value of
$\Lambda^2$ due to uncertainty in the details of the
$pp$ interaction.  The exponential potential yields a
central value of $\Lambda^2=6.960$.

Next, we determine the spread in the values of $\Lambda^2$ caused by
different deuteron potentials (using the exponential $pp$
potential).  The results are listed in Table II and range from
$\Lambda^2=6.917$ (for the RSC potential) to $\Lambda^2=6.988$
(for the Argonne v18 potential).  We also assign an uncertainty
of $\pm0.5\%$ to the value of $\Lambda^2$ due to uncertainty in the
deuteron wave function.  The Argonne v14 potential yields the
central value of $\Lambda^2=6.960$.

In the column labeled $\Lambda_{num}^2$ in Table 2, we list the
values of $\Lambda^2$ obtained by naively inserting the various
tabulated deuteron wave functions into the overlap integral.
These values vary from 6.915 to 7.129, which would naively imply an
uncertainty of about $\pm1.5\%$.  However, when the values of $\Lambda^2$
are corrected for the differences in the assumed values of $E_d$,
$\rho_d$, and $\eta_d$,
the resulting total uncertainty is only $\pm0.5\%$.  The
magnitude of the uncertainty in $\Lambda^2$ due to uncertainty
in the shape of the deuteron potential is similar to that due to
uncertainty in the shape of the $pp$ potential.

The differences in the various
published values for $\Lambda^2$ (see Table II in Bahcall and
Pinsonneault 1992) are primarily the result of
differences in the input parameters.  The
theoretical uncertainty is smaller than one would
estimate by simply comparing the published results.  For
example, the result given here is almost 2\% smaller than that obtained by
Bahcall and May (1969).  This difference can be traced
primarily to the high value of $E_d$ embodied in the deuteron
wave function used therein (from McGee 1966, who adopted
$E_d=2.267$~MeV), and
secondarily to smaller differences in the input $\eta_d$, $\rho_d$,
$\rho_p$, and $a_p$.  If we correct for these discrepancies in
input data using the
effective-range approximation, the Bahcall-May result
becomes 7.02, within 1\% of the current result.

The anomalously high value of $\Lambda^2=7.39$ obtained by Gould
and Guessoum (1990; Guessoum 1988) deserves special comment.
The authors state that this result was obtained using the
overlap of the Paris deuteron wave
function (Lacombe \etal\ 1980) and a $pp$ wave function
employing a Bargmann nuclear potential (see Noyes 1967) modified
to include a soft repulsive core.  From the discussion in
the papers by Gould and Guessoum, it not clear what
form they assumed for the $pp$ potential, nor is it clear what
values of  $a_p$ and $\rho_p$ they adopted.  Therefore, it is
difficult to make an unambiguous comparison with their result or
even to
understand precisely the origin of the difference between the
Gould-Guessoum value and all the other values given in Table 1
and Table 2 (as well as the other values obtained by different
authors; see Table II of Bahcall and Pinsonneault 1992).  We
have, however, carried out an illustrative calculation using a
potential that seems to be suggested by the description in Gould
and Guessoum.  We obtain a value of 7.42 for $\Lambda^2$,
consistent with the Gould-Guessoum value, by adding to our
exponential potential an attractive square-well core of depth 30
MeV and range 0.4 fm.  However, for this potential, we find
values of $a_p=-8.4826$ fm and $\rho_p=2.727$ fm, which are
inconsistent with the experimental data (discrepancies
$>200\sigma$ and $4\sigma$ for $a_p$ and $\rho_p$,
respectively).  We do not claim that this procedure was used by
Gould and Guessoum, but it is the only procedure we could think
of that reproduces their result.  In all of the numerical
experiments that we have
performed using wave functions and potentials that are
consistent with the current experimental data, we have never
obtained a value for $\Lambda^2$ greater than 7.00.  We conclude
that the Gould-Guessoum  value can only be obtained by using
some input data or some assumption that contradicts the existing
experimental information on the $pp$ system.

\bigskip
\centerline{6. VACUUM POLARIZATION}
\medskip

We next consider, following Bohannon and Heller (1977) and Gould
(1991), the effect of vacuum polarization (VP) on the
$pp$ wave function. In quantum electrodynamics, the
Coulomb potential is obtained from the Fourier transform of the
matrix element for scattering via exchange of one virtual photon
{}from an electrostatic source.  To next order in $\alpha$, the
photon propagator is augmented by an electron-positron loop.
This augmentation introduces an $\order(\alpha)$ correction to
the matrix
element which, when Fourier transformed, leads to a small correction,
the Uehling potential (Uehling, 1935), to the electrostatic potential.
The complete electrostatic potential becomes, in this approximation,
$$
{e^2\over r}+ {e^2\over r}\left({2\alpha I(r)\over
3\pi}\right),
\eqn\vpschrod
$$
where
$$
I(r)=\int_1^\infty\, e^{-2m_e rx} \left(1+{1\over 2x^2}\right)
{(x^2-1)^{1/2} \over x^2}\,dx.
\eqn\Ieqn
$$
The function $I(r)$ has the limiting forms
$$
I(r) = -\gamma - 5/6 - \ln(m_e r), \qquad {\rm for} \qquad m_e r\ll1,
\eqn\Ieqnone
$$
and
$$
I(r) = {3(2\pi)^{1/2} \over 4} {e^{-2 m_e r}\over (2 m_e r)^{3/2}}, \qquad
{\rm for} \qquad m_e r\gg1.
\eqn\Ieqntwo
$$
The function $I(r)$ has a logarithmic singularity for very small
radii, is of order a few until a radius of about $1/2m_e=193.1$
fm, and
then suffers an exponential falloff (arising from the exchange
of a virtual electron-positron pair) for $r\gap1/2m_e$.

A self-consistent determination of the effect of VP
on the calculation of $\Lambda^2$ must take into account the following
two effects.  First, VP must be incorporated in the
analysis of low-energy $pp$ scattering data, which
alters the inferred parameters in the nuclear potential.  Second, the
$pp$ wave function must be calculated with the Uehling potential in the
Schrodinger equation.  Using a sophisticated effective-range formalism
for vacuum polarization, Bohannon and Heller (1977) find that
including VP in the potential in the Schrodinger equation
decreases the $pp$ reaction rate by between 0.8\% and 1.2\%,
and using the WKB approximation, Gould (1990) finds that VP
decreases the $pp$ reaction rate by about 1.3\%.  However,
in both of these papers, the quoted result includes only the
second effect described above, \ie, the inclusion of
the Uehling
potential in the calculation of the $pp$ wave function.
Neither paper includes a calculation of the effect of VP on the
low-energy scattering parameters from which the nuclear
potential is determined.

We perform a numerical calculation of the
effect of VP on the $pp$ reaction rate which allows us to
isolate both effects of VP.  First, we consider the effect of VP on
the low-energy $pp$ scattering parameters, $a_p$ and $\rho_p$,
that are obtained experimentally.  These parameters are fit to
phase shifts measured at energies of roughly 0.3 MeV to 30 MeV.
The effective-range approximation
converges only for inverse wavenumbers $k^{-1}$ small compared
with the range of the nuclear potential.  Since the range of the
Uehling potential is $\sim200$ fm, we cannot estimate the effect
of the Uehling potential on $a_p$ and $\rho_p$ by using
Eqs.~\apeqn\ and \roeqn, and incorporating the Uehling potential
together with the nuclear potential.  We must calculate the
effect of VP on the phase shifts for the energies at which the
measurements are performed.

For a given nuclear potential, the phase
shifts are determined by integrating the Schrodinger equation,
Eq.~\schrod.  To account for VP, we make the
substitution
$$
V(r)\rightarrow V(r)\left[1+{2\alpha I(r) \over 3\pi} \right]
\eqn\subst
$$
in Eq.~\schrod.  We call the phase shifts obtained from this
substitution $\delta_{VP}(k)$, and we call the phase shifts
obtained from Eq.~\schrod\ ({\it without} VP) $\delta(k)$.
In fact, the measured phase shifts {\it are} $\delta_{VP}$, and
the low-energy scattering parameters are correctly obtained by
fitting to the relation
$$
C_0^2 k \cot\delta_{VP}(k)+{1\over R} h(kR) = -{1\over a_p} +
{1\over 2} \rho_p k^2,
\eqn\really
$$
instead of to Eq.~\expansion.  The $pp$ potentials used in
Section III were constructed using scattering parameters
obtained from Eq.~\expansion, not Eq.~\really.
Therefore, in constructing the $pp$ nuclear
potential as in Section III, we make the substitutions
$a_p\rightarrow a_p-\delta a_p$ and
$\rho_p \rightarrow \rho_p -\delta \rho_p$ in the low-energy
parameters, where $\delta a_p$ and $\delta \rho_p$ are obtained
by fitting the measured phase shifts to
$$
C_0^2 \cot\delta_{VP}+{1\over R} h(kR) - f(k) = -{1 \over a_p - \delta
a_p} + {1\over 2} (\rho_p-\delta \rho_p) k^2.
\eqn\deltaeqn
$$
Here, the function
$$
f(k)\equiv C_0^2 k [\cot\delta_{VP}(k) - \cot \delta(k)]
\eqn\deltaeqn
$$
is obtained for each data point by integrating the
Schrodinger equation numerically with and without the Uehling
potential using a nuclear potential that gives the correct $a_p$
and $\rho_p$.  We used an exponential nuclear potential, but the
results are insensitive (to an accuracy of about 0.1\% in $a_p$
and $\rho_p$) to the choice of potential.

We first calculate $\delta a_p$ and $\delta\rho_p$ with the aid
of a set of measured phase shifts given by Jackson
and Blatt (1950).  Although the modern data set consists of more
data points, the Jackson-Blatt data are sufficient (see
discussion below of results using part of the data and using
simulated data)
to evaluate the small effect of VP on the scattering parameters.
By first fitting the complete data set to Eq.~\really, and then
to Eq.~\deltaeqn, we find $\delta a_p=0.0687$ and $\delta \rho_p
= -0.030$.  Using Eq.~\effectiverange, we find that this results
in an increase in $\Lambda^2$ of 0.7\%.  The sign of this result
is expected:  If VP (a repulsive potential) is included, the
nuclear potential must be deeper to compensate, so $\Lambda^2$
is increased.  To assess the
dependence of our results on the data set used, we do the same
with only the lower-energy data points.  (The complete data set
consists of 24 data points at center-of-mass energies of
0.1765-3.53 MeV obtained with Van de Graaff generators, and five
data points with energies of 4.2-14.5 MeV obtained with
cyclotrons.  In their analysis, Jackson and Blatt disregarded the data
points obtained with cyclotrons since the data were less reliable at the
time.)  Using only the lower-energy data, we find $\delta
a_p=0.0732$ and $\delta \rho_p=-0.042$, which leads to an 0.4\%
increase in $\Lambda^2$.

We used simulated data to estimate the effect of
VP on the modern data set, which include measurements up to 30
MeV.  We calculated the effect of VP on
the values of $a_p$ and $\rho_p$ obtained from ten simulated
data points that exactly reproduce the measured $a_p$ and
$\rho_p$ in Eq.~\expansion\ which are uniformly spaced between 3
and 30 MeV.  This calculation gives
the correct fractional changes due to VP if the changes
are small.  For this simulated modern data set,
$\delta a_p=0.0321$
and $\delta\rho_p=-0.001$, which again results in an increase in
$\Lambda^2$ of 0.4\%.  Since this data set most-closely resembles
the modern data set, we choose 0.4\% to be the best estimate of the
correction to $\Lambda^2$ due to VP-corrections to the
low-energy scattering parameters.   To be
conservative, we estimate that the total theoretical uncertainty
includes the full range inferred above using the Jackson-Blatt
data and also includes a small uncertainty due to the choice of
the nuclear potential.  We conclude that correcting the low-energy
scattering parameters for VP results in an $0.4^{+0.4}_{-0.1}$\%
increase in $\Lambda^2$.

Next we evaluate the effect of including VP on the
wave function obtained by integrating the Schrodinger equation.
By numerically integrating
Eq.~\schrod\ with and without the Uehling potential, but using the
{\it same} nuclear potential, we find that the value of
$\Lambda^2$ is decreased by 0.9\% when the Uehling potential is
included.  The sensitivity of this
result to the choice of the shape of the nuclear potential is
less than 0.1\%.  Again, the sign of the effect is expected
since the Uehling potential is repulsive.  This is the result
for the correction to $\Lambda^2(0)$, the matrix element squared
at zero energy.  The most probable energy of interaction in the
Sun is about 6 keV (Bahcall 1989).  We use the WKB approximation
to evaluate the energy dependence of the VP correction to
$\Lambda^2$ (Gould 1990; Kamionkowski and Bahcall 1993).
The magnitude of this VP correction is about 10\% larger at 6
keV than at zero-energy, so we conclude that inclusion of the
Uehling potential in the calculation of the
wave function decreases $\Lambda^2$ by 1.0\% at solar energies.  Our
result is consistent with that of Bohannon and Heller, but
is slightly smaller than Gould's result.  This small difference
is most likely due to Gould's use of the WKB
approximation (see Kamionkowski and Bahcall 1993).

Combining our results from including vacuum polarization in
analyzing the scattering data and in integrating Schrodinger's
equation, we find that VP decreases $\Lambda^2$ by
$0.6^{+0.1}_{-0.4}$\%.  The {\it net} effect we find for VP on the $pp$
matrix element is about half that found by
Bohannon and Heller (1977) and Gould (1991).  We attribute the
difference to the fact that we self-consistently include the effect of VP
on the measured low-energy scattering parameters, whereas this aspect of
the influence of VP (which partially cancels the effect of VP in
the potential used in integrating the Schrodinger equation) was
not considered by Gould and was not isolated by Bohannon and Heller.

\bigskip
\centerline{7. SUMMARY AND DISCUSSION}
\medskip

We have calculated the matrix element for the
reaction $p+p\rar {}^2{\rm H} +e^+ +\nu_e$ with and without
the effects of vacuum polarization.
Without vacuum
polarization, we find the square of the overlap integral to be
$\Lambda^2= 6.96\times(1\pm0.002\pm0.010\%)$, where the first
uncertainty results from $1\sigma$ errors in the experimental
quantities and the second reflects imprecise knowledge of the
shape of the nuclear potential.
We include vacuum polarization self-consistently in the low-energy
$pp$ scattering parameters and in the numerical
calculation of the $pp$ wave function.
We find that vacuum polarization decreases $\Lambda^2$ by
$0.6^{+0.1}_{-0.4}$\%.  Our final
result is $\Lambda^2= 6.92\times(1\pm0.002^{+0.014}_{-0.009})$.
We show elsewhere that vacuum polarization will decrease the
rates for all the other
reactions in the $pp$ chain and the CNO cycle by small amounts,
less than 5\% (Kamionkowski and Bahcall 1993).

In addition to the $\order(\alpha)$ VP corrections to the $pp$
potential, there are $\order(\alpha)$ radiative corrections
(involving an extra soft photon in the final state) to the $pp$
reaction.  One can use the similarity of the radiative corrections to
the axial-vector part of neutron decay to those for proton decay
in the $pp$ reaction to account for these radiative corrections
to the $pp$ reaction (Bahcall and May 1968).  This correction
has been included in Eq.~\Szero.

It is conventional to use the low-energy cross-section factor,
$S(0)$, in stellar-evolution calculations.  Following the
discussion in Bahcall \etal\ (1982) and Bahcall and Pinsonneault
(1992), we find [\cf\ Eq.~\Szero]:
$$
S_{pp}(0)=3.89\times10^{-25}\,(1\pm0.011)\, {\rm MeV-barns}.
\eqn\Sppeqn
$$
In calculating the error given in Eq.~\Sppeqn, we have used
$1\sigma$ errors for experimentally-measured quantities (such as
$G_A/G_V$ or $a_p$) and one-third the {\it total} range of
values for theoretically-calculated effects (such as the
uncertainty from nuclear potentials or from mesonic
corrections).  Although the theoretical errors cannot be used in
a precise statistical sense, the uncertainty quoted in Eq.~\Sppeqn\ is
intended to be used as an approximate $1\sigma$ error in Monte
Carlo studies (\cf\ Bahcall and Ulrich 1988) of the overall
uncertainty in predictions of solar-neutrino event rates.  Our
best estimate is about 3\% smaller than that quoted by Bahcall
and Pinsonneault.  This is due in part (2\%) to our improved
value for $\Lambda^2$, and in part (1\%) to an updated value for
$G_A/G_V$ (Hikasa \etal\ 1992).

Our results imply a 7.5\% increase in the predicted event rates
for the Kamiokande (Hirata \etal\
1991) solar-neutrino experiment and a 6\% (0.5 SNU) increase for
the chlorine solar-neutrino experiment relative to the calculations of
Bahcall and Pinsonneault (1992), slightly increasing the
discrepancy between standard-model predictions and observations.
The predicted event rate for the Borexino experiment (sensitive
to $^7$Be, Raghavan 1990) is increased by about 3\%.

\bigskip

It is a pleasure to thank F.~Dyson, S.~Freedman, R.~Gould,
W.~Haxton, L.~Heller, F.~Wilczek, and R.~Wiringa
for valuable conversations and suggestions.  We are especially
grateful to R.~Gould for detailed, constructive comments on a
draft version of the manuscript.  MK was supported by
the Texas National Laboratory Research Commission, and by the
DOE through Grant No. DE-FG02-90ER40542.  JNB was supported by
the NSF through Grant No. PHY92-45317.

\endpage

\centerline{REFERENCES}
\medskip
\Rf{Bahcall, J. N. 1989, {\it Neutrino Astrophysics}\ (Cambridge
University Press, Cambridge).}
\Rf{Bahcall, J. N., and May, R. M. 1968, {\sl Ap. J. (Letters)},
{\bf 152}, L17.}
\Rf{Bahcall J. N., and May, R. M. 1969, {\sl Ap. J.} {\bf
155}, \rm 501.}
\Rf{Bahcall J. N. \etal\ 1982, {\sl Rev. Mod. Phys.} {\bf
54}, 767.}
\Rf{Bahcall J. N., and Ulrich, R. K. 1988, {\sl Rev. Mod. Phys.} {\bf
60}, 297.}
\Rf{Bahcall, J. N., and Pinsonneault, M. H. 1992, {\sl Rev. Mod.
Phys.} {\bf 64}, 885.}
\Rf{Bargholtz, C. 1979, {\sl Ap. J. (Letters)} {\bf 233}, L161.}
\Rf{Bergervoet J. R. \etal\ 1988, {\sl Phys. Rev. C} {\bf
38}, 15.}
\Rf{Bethe, H., and Critchfield, C. L. 1938, {\sl Phys. Rev.} {\bf
54}, 248.}
\Rf{Blin-Stoyle, R. J., and Papageorgiou, S. 1965, {\sl Nucl.
Phys.} {\bf 64}, 1.
\Rf{Bohannon, G. E., and Heller, L. 1977, {\sl Phys. Rev. C}
{\bf 15}, 1221.}
\Rf{Brandenburg, R. A. \etal\ 1988, {\sl Phys. Rev. C}
{\bf 37}, 1245.}
\Rf{Brolley, J. E. 1971, {\sl Sol. Phys.} {\bf 20}, 249.}
\Rf{Carlson, J., Riska, D. O., Schiavilla, R., and
Wiringa, R. B. 1991, {\sl Phys. Rev. C} {\bf 44}, 619.}
\Rf{Dautry, F., Rho, M., and Riska, D. O. 1976, {\sl Nucl.
Phys.} {\bf A264}, 507.}
\Rf{Davis, R., Jr. 1989, in {\it Proceedings of the Thirteenth
International Conference on Neutrino Physics and Astrophysics},
Boston, Massachusetts, 5-11 June 1988, edited by J. Schneps
\etal\ (World Scientific, Singapore), p. 518.}
\Rf{de Tourreil, R., and Sprung, D. W. L. 1973 {\sl Nucl. Phys.}
{\bf A201}, 193.}
\Rf{Ellis, S. D., and Bahcall, J. N. 1968, {\sl Nucl. Phys.} {\bf
A114}, 636.}
\Rf{Gari, M. 1978, in {\it Proceedings of Informal Conference on
Status and Future of Solar Neutrino Research}, edited by G.
Friedlander (Brookhaven National Laboratory), Report No.
50879, Vol.~1, p.~137.}
\Rf{Gari, M., and Huffman, A. H. 1972, {\sl Ap. J.} {\bf
178}, 543.}
\Rf{Gould, R. J. 1990 {\sl Ap. J.} {\bf 363}, 574.}
\Rf{Gould, R. J., and Guessoum, N. 1990, {\sl Ap. J. (Letters)}
{\bf 359}, L67.}
\Rf{Guessoum, N. 1988, Ph.D Thesis, University of
California, San Diego.}
\Rf{Hirata, K. S. \etal\ 1991, {\sl Phys. Rev. D} {\bf 44}, 2241.}
\Rf{Hikasa, K. \etal\ 1992, {\sl Phys. Rev. D} {\bf 45}, S1.}
1.}
\Rf{Jackson J. D., and Blatt, J. M. 1950 {\sl Rev. Mod.
Phys.} {\bf 22}, 77.}
\Rf{Kamionkowski, M., and Bahcall, J. N. 1993, IASSNS-AST-93/21.}
\Rf{Lacombe, M. \etal\ 1980, {\sl Phys. Rev. C} {\bf 21}, 861.}
\Rf{Lagaris, I. E., and Pandharipande, V. R. 1981, {\sl Nucl.
Phys.} {\bf A359}, 331.}
\Rf{Noyes, H. P. 1968, in {\it Few-Body Problems, Light Nuclei,
and Nuclear Interactions}, Vol. I, ed. G. Paic and J. Slaus (New
York, Gordon and Breach), p. 9.}
\Rf{Raghavan, R. S. 1990, in {\it Proceedings of the 25th
International Conference on High Energy Physics}, edited by K.
Phu and Y. Yamaguchi (World Scientific, Singapore), Vol. I, p.
482.}
\Rf{Reid, R. V. 1968, {\sl Ann. Phys.} {\bf 50}, 411.}
\Rf{Salpeter, E. E. 1952, {\sl Phys. Rev.} {\bf 88}, 547.}
\Rf{Uehling, E. A. 1935, {\sl Phys. Rev.} {\bf 48}, 55.}
\Rf{Wiringa, R. B., Smith, R. A., and Ainsworth, T. L. 1984,
{\sl Phys. Rev. C} {\bf 29}, 1207.}
\Rf{Wiringa, R. B. 1993, work in progress.}

\endpage

\figout
\endpage

\vbox{\tabskip=0pt \offinterlineskip
\def\tablerule{\noalign{\hrule}}
\halign to450pt{\strut#& \vrule#\tabskip=1em plus2em&
  \hfil#& \vrule#& \hfil#& \vrule#& \hfil#& \vrule#& \hfil#&\vrule#
  \tabskip=0pt\cr\tablerule
&&\multispan7\hfil Proton-Proton Potentials\hfil&\cr\tablerule
&&\omit\hidewidth Potential\hidewidth&&
 \omit\hidewidth $V_0$ (MeV)\hidewidth&& \omit\hidewidth $b$ (fm)\hidewidth&&
 \omit\hidewidth $\Lambda^2$\hidewidth&\cr\tablerule
&&\hfil SW\hfil&&-11.751&&2.7718&&6.916&\cr\tablerule
&&\hfil Gaussian\hfil&&-27.729&&1.8912&&6.937&\cr\tablerule
&&\hfil Exponential\hfil&&-98.861&&0.7407&&6.960&\cr\tablerule
&&\hfil Yukawa\hfil&&-46.124&&1.1809&&6.979&\cr\tablerule
&&\hfil RC\hfil&&-314.704&&0.5565&&6.939&\cr\tablerule
}}
\item{\rm Table~1.}  Proton-Proton potential parameters.  The values
of $\Lambda^2$ were obtained using the Argonne v14 deuteron
potential.
\vskip2cm

\vbox{\tabskip=0pt \offinterlineskip
\def\tablerule{\noalign{\hrule}}
\halign to450pt{\strut#& \vrule#\tabskip=1em plus2em&
  \hfil#& \vrule#& \hfil#& \vrule#& \hfil#& \vrule#&
  \hfil#&\vrule#& \hfil#&\vrule#& \hfil#&\vrule#& \hfil#&\vrule#
  \tabskip=0pt\cr\tablerule
&&\multispan{13}\hfil Deuteron Potentials\hfil&\cr\tablerule
&&\omit\hidewidth Potential\hidewidth&&
 \omit\hidewidth $E_d$ (MeV)\hidewidth&& \omit\hidewidth $\eta_d$\hidewidth&&
 \omit\hidewidth $\rho_d$ (fm)\hidewidth&&
 \omit\hidewidth $\Lambda_{num}^2$\hidewidth&&
 \omit\hidewidth $\Lambda_{eff}^2$\hidewidth&&
 \omit\hidewidth $\Lambda^2$\hidewidth&\cr\tablerule
&&\hfil Experimental\hfil&&
    2.224575&&0.0256&&1.759&&  &&6.975&& &\cr\tablerule
&&\hfil SSC\hfil&&
    2.224066&&0.0255&&1.833&&7.129&&7.128&&6.977&\cr\tablerule
&&\hfil Urbana v14\hfil&&
    2.224637&&0.0254&&1.816&&7.099&&7.092&&6.981&\cr\tablerule
&&\hfil Argonne v18\hfil&&
    2.224575&&0.0253&&1.770&&7.009&&7.000&&6.988&\cr\tablerule
&&\hfil Argonne v14\hfil&&
    2.224884&&0.0266&&1.805&&7.055&&7.070&&6.960&\cr\tablerule
&&\hfil RSC\hfil&&
    2.224688&&0.0262&&1.758&&6.915&&6.973&&6.917&\cr\tablerule
&&\hfil McGee\hfil&&
    2.2669104&&0.0269&&1.749&&7.026&&7.031&&6.970&\cr\tablerule
}}
\item{\rm Table~2.}  Deuteron potentials.  The values of $\Lambda^2$
were obtained using the exponential proton-proton potential.

\vskip2cm
\vfill
\eject

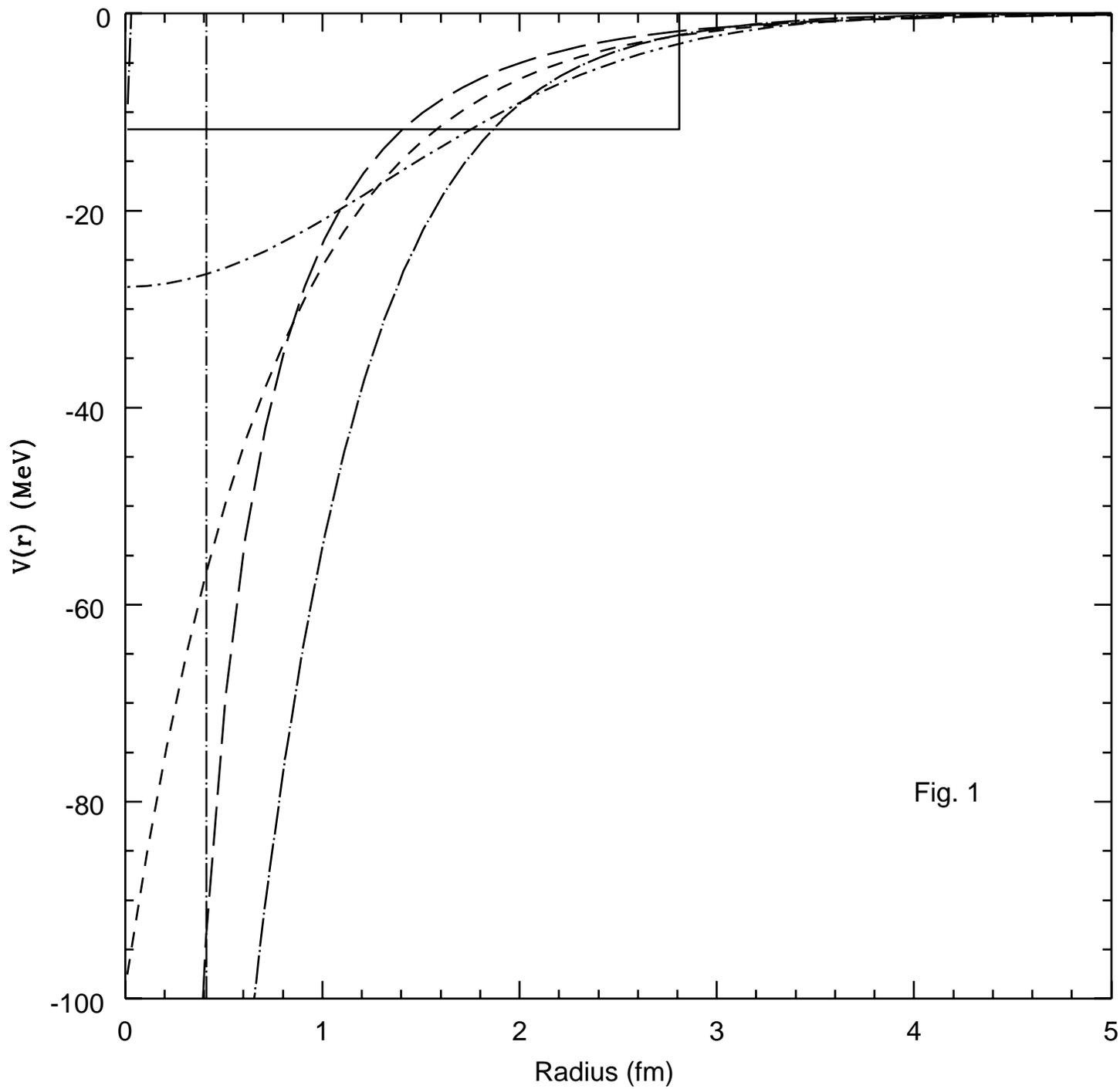
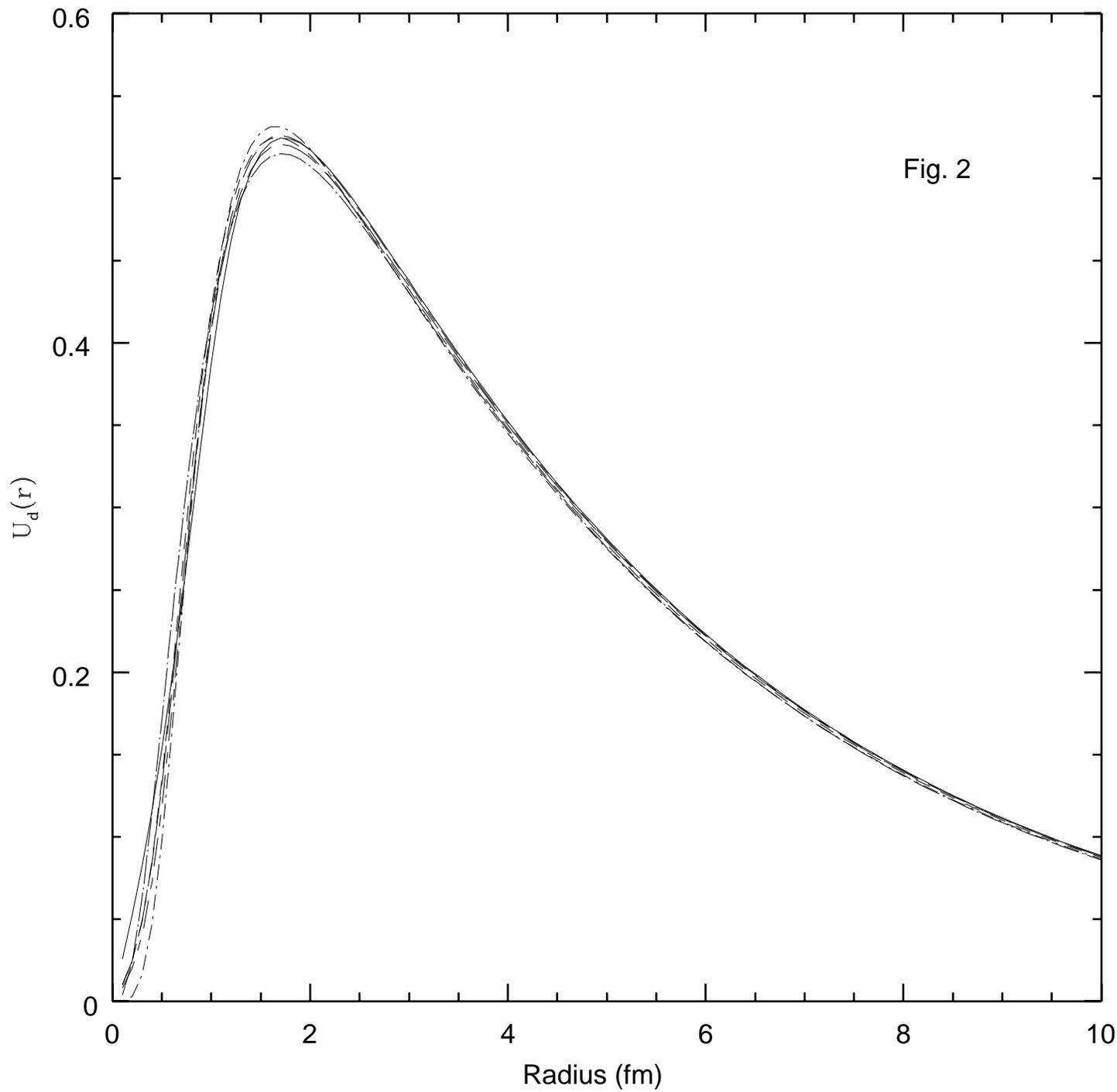
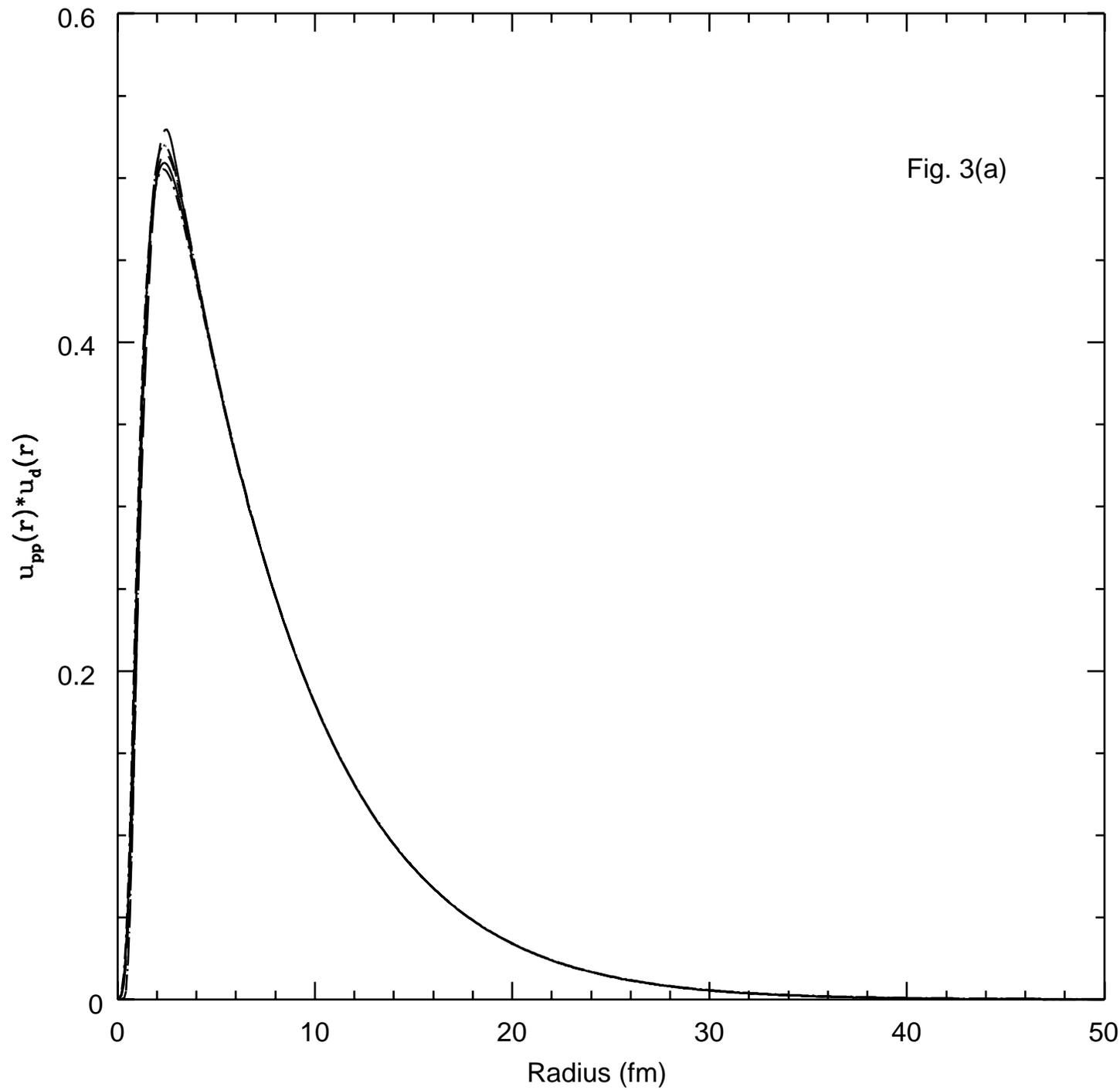
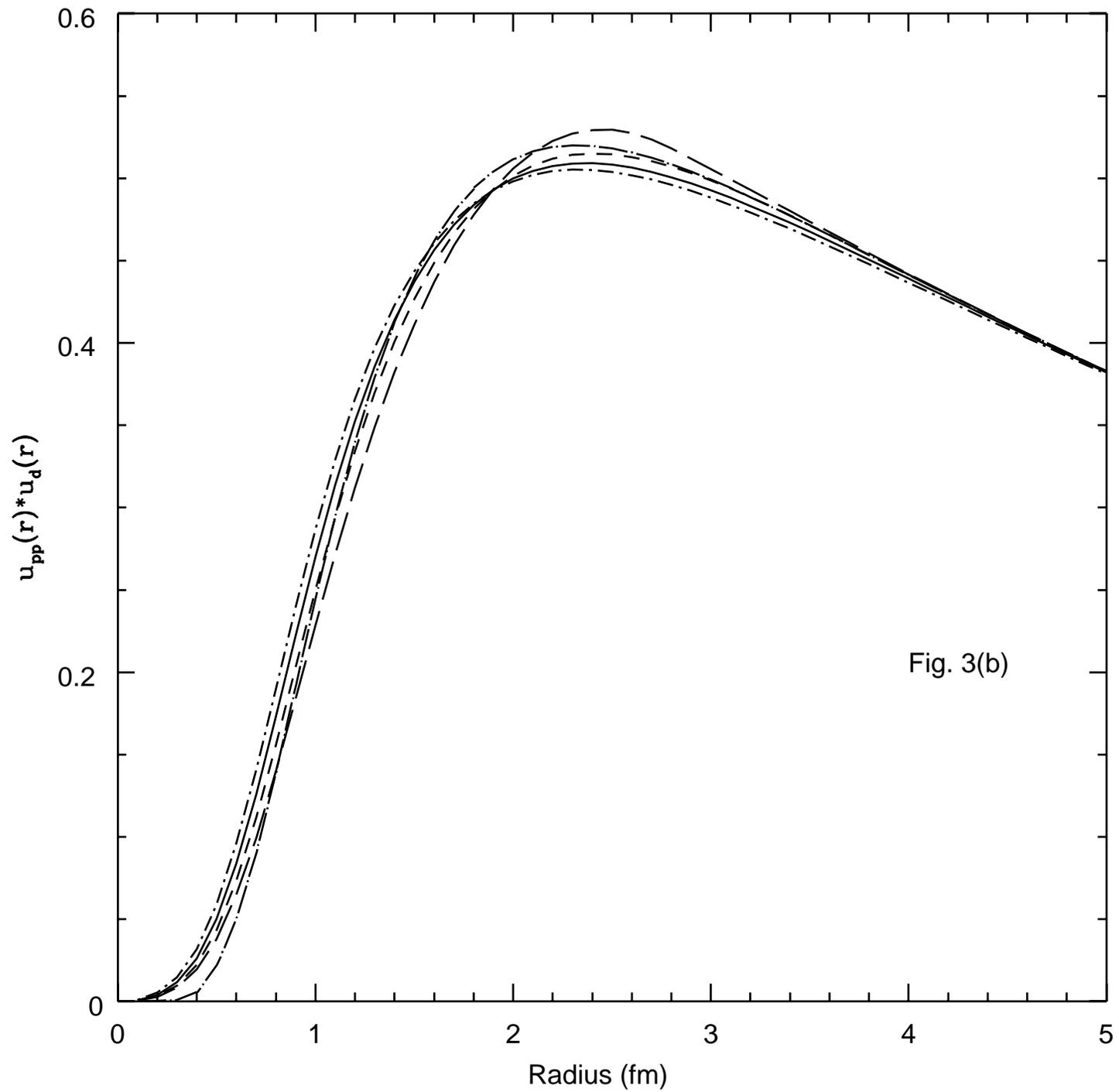
\end